\documentstyle[12pt,epsfig]{article}
\newcommand{\be}{\begin{eqnarray}}
\newcommand{\ee}{\end{eqnarray}}

\newcommand{\lam}{\lambda}

\begin{document}
\title{Moment--operator expansion for the two-meson, two-photon and
fermion-antifermion states}
\author{A.V. Anisovich, V.V. Anisovich, V.N. Markov,\\
M.A. Matveev and A.V. Sarantsev\\
{\it St. Petersburg Nuclear Physics Institute, 188300, Gatchina,
Russia}}
\date{}
\maketitle

\begin{abstract}
A complete set of formulae in terms of the moment--operator expansion
is presented for the  two-meson, two-photon and
fermion-antifermion states
that can be used in the
analysis of scalar $(S)$ and
pseudoscalar $(P)$ meson production in $p\bar p$ and $\gamma\gamma$
collisions: $p\bar p \to SS,PP,SP$ and $\gamma\gamma\to SS,PP,SP$.
Method of a generalization of the formulae for amplitudes of
multi-meson production in the two-stage processes of the type  {\it
(spin-$j$) resonance $+$ meson $\to$ three mesons} is also discussed.
\end{abstract}

\section{Introduction}

The moment--operator expansion is a powerful
tool for the study of the analytical structure of
amplitudes,  in particular, for the
determination of the resonance amplitudes.

In this paper we present a full set of formulae which are necessary for
the analysis of reactions $p\bar p\to\pi\pi$, $p\bar p\to\eta\eta$,
$p\bar p\to\eta\eta'$, $p\bar p\to\pi f_0$ in flight,
and we explain the  way of generalization of the formulae for
the analysis of reactions with the production of
resonances with non-zero spin,
 such as $\pi \rho$, $\pi f_2$, $\pi \pi_2$, and so on.

The data obtained
by Crystal Barrel Collaboration provide us with rich possibilities to
study  resonances in the mass region
1900-2400 MeV \cite{anis-1,anis-2}. 
The reconstruction of analytical meson amplitudes and definition
of pole singularities are forced by the problem of classification of
meson states, exotics included (glueballs, hybrids). The establishing
of a complete set of meson states is a necessary step in understanding
of Strong QCD, see e.g. \cite{klempt,montanet,petry,AAS}.

We give formulae for
$\gamma\gamma$-states; our interest in these processes is stimulated by
rich experimental information on $\gamma\gamma\to hadrons$ obtained at
LEP (e.g. see  \cite{Kienzle,Scheg}).

 Starting from early 1960s, the operator expansion was
exploited for the amplitude analysis within the framework of
dispersion--relation technique:  in \cite{AD} the operator expansion
was used for the calculation of rescattering processes in the reaction
of three-particle production near threshold.  A complete description of
the method for non-relativistic kinematics was presented in
\cite{Zemach}.

The moment-operator expansion method  is especially helpful for the
investigation of particle spectra in the
framework of multichannel $K$-matrix
approach, or dispersion--relation $N/D$ method.
Within this method, the calculation of rescattering effects in
multiparticle reactions is rather simple and straightforward.

For the two-fermion sector of nucleon-nucleon reactions, such as
$NN\to NN$, $NN\to N\Delta$ and $\gamma d\to pn$, the relativistic
operators have been constructed in \cite{NN,AKMS,N-Delta,gamma-d}. In
\cite{3pi} the operator expansion has been carried out for the reaction
$p\bar p \to \pi\pi\pi$, and it was extended for the coupled-channel
reactions $p\bar p\to\pi\pi\pi$, $p\bar p\to\eta\eta\pi$,
$p\bar p\to K\bar K\pi$ in \cite{exp-3pi}, within the framework of the
$N/D$ method.

One of the advantages of the operator expansion method
is a possibility to take promptly into accont
kinematical factors related to momenta of the incoming and
outgoing particles.
The account for kinematical factors is dictated by the necessity to
investigate analytical structure of the amplitude near the production
threshold  of new particles. An alternative to the operator-expansion
method consists in calculating kinematical factors in a standard
expansion in spherical harmonics (see \cite{Chung} as an example).

The moment-operator expansion has been recently used in the calculation
of the radiative decays of scalar, tensor and vector mesons in the
framework of  double spectral integration technique \cite{f-gg,AA,phi}.

First of all, we present here the angular--momentum operator
$X^{(J)}_{\mu_1\mu_2\ldots\mu_{J-1}\mu_J}(k)$ for the two spinless
mesons (Section 2). In terms of these
operators we construct the scattering amplitudes for the processes
{\it two mesons $\to$ two mesons} and compare it with an
amplitude for three-meson production in the  two-stage process: {\it two
mesons $\to$ (spin-$j$ resonance)+meson $\to$ three
mesons}.  By comparing these two amplitudes one can clearly see the
difference between our  and Zemach's \cite{Zemach} approaches  (or the
approach with spherical harmonics, see e.g. \cite{Chung}). Indeed, the
description of the transition {\it two mesons $\to$ three mesons}
in the approaches \cite{Zemach,Chung} requires additional Lorentz
boost, which is different for various regions of the three-particle
Dalitz-plot, while one does not need such a boost in
the relativistic covariant method presented here.

The angular--momentum operator for the two-fermion system,
$Q^{(S,L,J)}_{\mu_1\mu_2\ldots\mu_{J-1}\mu_J}$, is constructed in
Section 3. We consider in detail the amplitude of
the reaction $p\bar p \to${\it  two spinless  mesons}.
Denote meson momenta as $k_1$, $k_2$ and fermion momenta as $q_1$,
$q_2$.  The amplitude with total angular momentum $J$ has the following
stucture:
\be
\bar\psi(-q_2)Q^{(S,L,J)}_{\mu_1\mu_2\ldots\mu_{J-1}\mu_J}\psi(q_1)
X^{(J)}_{\mu_1\mu_2\ldots\mu_{J-1}\mu_J}(k)A_{S,L,J}(s).
\label{2.15}
\ee
The operator
$\bar\psi(-q_2)Q^{(S,L,J)}_{\mu_1\mu_2\ldots\mu_{J-1}\mu_J}\psi(q_1)$
refers to the $p\bar p$ state: here
$S$ is the total spin of fermions, $S=0,1$, and $L$ is
the angular momentum. The function
$X^{(J)}_{\mu_1\mu_2\ldots\mu_{J-1}\mu_J}(k)$ stands for
the two-meson state operator: it is the angular--momentum operator
which depends on relative momentum of mesons, $k=(k_1-k_2)/2$.
Partial--wave amplitude $A_{S,L,J}(s)$ is a function of the invariant
energy squared, $s=P^2$, where $P=k_1+k_2=q_1+q_2$.

To be illustrative, in Section 4 we present formulae which
are helpful for the study of the reactions of $p\bar p$ annihilation,
with the  production of two spinless particles, $p\bar p \to SS, \; PP,
\; SP$: these formulae were used in the analyses  \cite{anis-1,anis-2}.

For $\gamma\gamma$ states, the spin and
angular--momentum operators are given in Section 5. We construct the even
and odd operators and demonstrate the connection of spin operators with
helicity amplitudes.

\section{Angular--momentum operator
$\; X^{(L)}_{\mu_1\mu_2\ldots\mu_{L-1}\mu_L}$  \\
and the two-meson scattering amplitude }

Here we construct the operator
$ X^{(L)}_{\mu_1\mu_2\ldots\mu_{L-1}\mu_L}(k)$  and describe
its properties. Then we present the partial--wave amplitude
in terms of the introduced operator expansion.

\subsection{Operator
$ X^{(L)}_{\mu_1\mu_2\ldots\mu_{L-1}\mu_L}(k)$ }

The operator $ X^{(L)}_{\mu_1\mu_2\ldots\mu_{L-1}\mu_L}(k)$
is constructed by using the relative momentum
of mesons in the space orthogonal to
the total momentum $P$:
\be
k^\perp_\mu = k_\nu
g^\perp_{\nu\mu},  \qquad
g^\perp_{\nu\mu}=g_{\nu\mu}-\frac{P_\nu
P_\mu}{s}.
\ee
In the
centre-of-mass system, where $P=(P_0,\vec P)=(\sqrt s,0)$, the vector
$k^\perp$ is space-like: $k^\perp=(0,\vec k)$.
We determine the operator
$ X^{(L)}_{\mu_1\mu_2\ldots\mu_{L-1}\mu_L}(k)$
as symmetrical and traceless. It is easy to construct it for the lowest
values of $L=0,1,2,3,4$:
\be
X^{(0)}=1\ , \qquad X^{(1)}_\mu=k^\perp_\mu\ , \qquad
X^{(2)}_{\mu_1 \mu_2}=\frac32\left(k^\perp_{\mu_1}
k^\perp_{\mu_2}-\frac13\, k^2_\perp g^\perp_{\mu_1\mu_2}\right),
\label{2.2}
\ee
$$
X^{(3)}_{\mu_1\mu_2\mu_3}=\frac52\left[k^\perp_{\mu_1} k^\perp_{\mu_2 }
k^\perp_{\mu_3}-\frac{k^2_\perp}5\left(g^\perp_{\mu_1\mu_2}k^\perp
_{\mu_3}+g^\perp_{\mu_1\mu_3}k^\perp_{\mu_2}+
g^\perp_{\mu_2\mu_3}k^\perp_{\mu_1}
\right)\right] ,
$$
$$
X^{(4)}_{\mu_1\mu_2\mu_3\mu_4}=\frac{35}{8}\left[k^\perp_{\mu_1}
k^\perp_{\mu_2 }k^\perp_{\mu_3}k^\perp_{\mu_4}
\right.
$$
$$
-\frac{k^2_\perp}7\left(
g^\perp_{\mu_1\mu_2}k^\perp_{\mu_3}k^\perp_{\mu_4}+
g^\perp_{\mu_1\mu_3}k^\perp_{\mu_2}k^\perp_{\mu_4}+
g^\perp_{\mu_1\mu_4}k^\perp_{\mu_2}k^\perp_{\mu_3}+
g^\perp_{\mu_2\mu_3}k^\perp_{\mu_1}k^\perp_{\mu_4}+
g^\perp_{\mu_2\mu_4}k^\perp_{\mu_1}k^\perp_{\mu_3}+
g^\perp_{\mu_3\mu_4}k^\perp_{\mu_1}k^\perp_{\mu_2}\right)
$$
$$
\left .
+\frac{k^4_\perp}{35}\left(
g^\perp_{\mu_1\mu_2}g^\perp_{\mu_3\mu_4}+
g^\perp_{\mu_1\mu_3}g^\perp_{\mu_2\mu_4}+
g^\perp_{\mu_1\mu_4}g^\perp_{\mu_2\mu_3}\right)
\right]
$$
Correspondingly, the generalization of
$X^{(L)}_{\mu_1\ldots\mu_L}$ for
$L>1$ reads:
\be
\label{2.3}
X^{(L)}_{\mu_1\ldots\mu_L}&=&k^\perp_\alpha
Z^{(L-1)}_{\mu_1\ldots\mu_L, \alpha} \  , \\
Z^{(L-1)}_{\mu_1\ldots\mu_L, \alpha}&=&
\frac{2L-1}{L^2}\left (
\sum^L_{i=1}X^{{(L-1)}}_{\mu_1\ldots\mu_{i-1}\mu_{i+1}\ldots\mu_L}
g^\perp_{\mu_i\alpha} \right .
\nonumber \\
&&\left . -\frac{2}{2L-1} \sum^L_{i,j=1 \atop i<j}
g^\perp_{\mu_i\mu_j}
X^{{(L-1)}}_{\mu_1\ldots\mu_{i-1}\mu_{i+1}\ldots\mu_{j-1}\mu_{j+1}
\ldots\mu_L\alpha} \right )\ .
\nonumber
\ee
It is seen that the operator
$ X^{(L)}_{\mu_1\mu_2\ldots\mu_{L-1}\mu_L}(k)$
constructed in accordance with (\ref{2.3})
is symmetrical,
\be
X^{(L)}_{\mu_1\ldots\mu_i\ldots\mu_j\ldots\mu_L}\; =\;
X^{(L)}_{\mu_1\ldots\mu_j\ldots\mu_i\ldots\mu_L},
\label{2.12}
\ee
and it works in the space orthogonal to $P$:
\be
P_{\mu_i}X^{(L)}_{\mu_1\ldots\mu_i\ldots\mu_L}\ =\ 0 \ .
\label{2.13}
\ee
The angular-momentum operator
$X^{(L)}_{\mu_1\ldots\mu_L}$ is traceless over any two
indices:
\be
g_{\mu_i\mu_j}X^{(L)} _{\mu_1\ldots\mu_i\ldots\mu_j\ldots\mu_L}\
=\ g^\perp _{\mu_i\mu_j}X^{(L)} _{\mu_1\ldots\mu_i\ldots\mu_j\ldots\mu_L}
\ =\ 0.
\label{2.1}
\ee
The tracelessness property given by
(\ref{2.1}) is obvious for the lowest--order operators entering
(\ref{2.2}), for example, $g^\perp_{\mu_1\mu_2}X^{(2)}_{\mu_1\mu_2}=0$
(recall that $g^\perp_{\mu_1\mu_2} g^\perp_{\mu_1\mu_2}=3$).  Now let
us demonstrate that, if $X^{{(L-1)}}_{\mu_1\ldots\mu_{L-1}}$ is
traceless, the operator $X^{(L)}_{\mu_1\ldots\mu_L}$ is traceless as
well.  The symmetry condition tells us that it would be enough to prove
$g^\perp_{\mu_1\mu_2}X^{(L)}_{\mu_1\mu_2\ldots\mu_L}=0$.
The latter  is directly seen from the definition (\ref{2.3}):
\begin{eqnarray*}
g^\perp_{\mu_1\mu_2} X^{(L)}_{\mu_1\mu_2\ldots\mu_L}&
=& k^\perp_\alpha\frac{2L-1}{L^2}\left(2X^{{(L-1)}}_{\mu_3 \ldots
\mu_L\alpha}-\frac2{2L-1}\left [
2(L-2)X^{{(L-1)}}_{\mu_3\ldots\mu_L\alpha}
+3X^{{(L-1)}}_{\mu_3\ldots\mu_L\alpha}\right ]\right)\ \\
&=& k^\perp_\alpha X^{{(L-1)}}_{\mu_3\ldots\mu_L\alpha}\
\frac{2L-1}{L^2} \left[2-\frac2{2L-1}\ (2L-1)\right] = \; 0 \ .
\end{eqnarray*}
The convolution equality reads:
\be
X^{(L)}_{\mu_1\ldots\mu_L}k^\perp_{\mu_L}=k^2_\perp
X^{(L-1)}_{\mu_1\ldots\mu_{L-1}}\ .
\label{2.4}
\ee
It can be easily proven by the induction method. Indeed, one has
for the lowest moments: $k^\perp_\mu X^{(1)}_\mu=k^2_\perp
X^{(0)}$.  Using
$X^{(L-1)}_{\mu_1\ldots\mu_{L-1}}k^\perp_{\mu_{L-1}}=k^2_\perp
X^{(L-2)}_{\mu_1\ldots\mu_{L-2}}$
as an input we obtain from (\ref{2.3}):
\be
&& X^{(L)}_{\mu_1\ldots\mu_L}k^\perp_{\mu_L}=
k^2_\perp \frac{2L-1}{L^2}
\bigg ( X^{{(L-1)}}_{\mu_1\ldots\mu_{L-1}} \frac{2L-3}{2L-1}\times
\bigg .
\nonumber \\
&&
\times\bigg [\sum^{L-1}_{i=1}
X^{(L-2)}_{\mu_1\ldots\mu_{i-1}\mu_{i+1}\ldots\mu_{L-1}}k^\perp_{\mu_i}-
\frac2{2L-3}\sum^{L-1}_{i,j=1 \atop i<j}g^\perp_{\mu_i\mu_j}
X^{(L-2)}_{\mu_1\ldots\mu_{i-1}\mu_{i+1}\ldots\mu_{j-1}\mu_{j+1}
\ldots\mu_{L-1}\alpha} k^\perp_{\alpha}
\bigg ] \bigg )
\nonumber \\
&&=k^2_\perp\frac{2L-1}{L^2}X^{{(L-1)}}_{\mu_1\ldots\mu_{L-1}}
\left[1+\frac{(L-1)^2}{2L-1}\right]=k^2_\perp
X^{\left( L-1\right)}_{\mu_1\ldots\mu_{L-1}} \ .
\nonumber
\ee
Using (\ref{2.4}) we rewrite the recurrent equation (\ref{2.3}) in the
form:
\be
X^{(L)}_{\mu_1\ldots\mu_L}= \frac{2L-1}{L^2}
\sum^L_{i=1}k^\perp_{\mu_i}
X^{{(L-1)}}_{\mu_1\ldots\mu_{i-1}\mu_{i+1}\ldots\mu_L}
-\frac{2k^2_\perp}{L^2}\sum^L_{i,j=1 \atop i<j}
g_{\mu_i\mu_j}^\perp
X^{\left(L-2\right)}_{\mu_1\ldots\mu_{i-1}\mu_{i+1}\ldots\mu_{j-1}\mu_{j+1}
\ldots\mu_L}\ .
\label{2.5}
\ee
Basing on this recurrent equation and
taking into account the tracelessness of $X^{(L)}_{\mu_1\ldots\mu_L}$,
one can write the normalization condition for the moment-$L$ operator
as follows:
\be
X^{(L)}_{\mu_1\ldots\mu_L}(k)X^{(L)}_{\mu_1\ldots\mu_L}(k)
=\alpha(L)k^{2L}_\perp \ ,
\qquad
\alpha(L)=\prod^L_{l=1}\frac{2l-1}{l}=\frac{(2L-1)!!}{L!} \ .
\label{2.10}
\ee
The iteration of equation (\ref{2.5}) gives us the
following expression for the operator $X^{(L)}_{\mu_1\ldots\mu_L}$:
\be
&&X^{(L)}_{\mu_1\ldots\mu_L}(k)
=\frac{(2L-1)!!}{L!}
\bigg [
k^\perp_{\mu_1}k^\perp_{\mu_2}k^\perp_{\mu_3}k^\perp_{\mu_4}
\ldots k^\perp_{\mu_L} -  \\
&&-\frac{k^2_\perp}{2L-1}\left(
g^\perp_{\mu_1\mu_2}k^\perp_{\mu_3}k^\perp_{\mu_4}\ldots k^\perp_{\mu_L}
+g^\perp_{\mu_1\mu_3}k^\perp_{\mu_2}k^\perp_{\mu_4}\ldots
k^\perp_{\mu_L} + \ldots \right)+
\nonumber
\\
&&
+\frac{k^4_\perp}{(2L-1)
(2L-3)}\left(
g^\perp_{\mu_1\mu_2}g^\perp_{\mu_3\mu_4}k^\perp_{\mu_5}
k^\perp_{\mu_6}\ldots k_{\mu_L}+
g^\perp_{\mu_1\mu_2}g^\perp_{\mu_3\mu_5}k^\perp_{\mu_4}
k^\perp_{\mu_6}\ldots k_{\mu_L}+
\ldots\right)+\ldots\bigg ]\ .
\nonumber
\label{x-direct}
\ee
Let us write one more useful equation,
\be
k_{\alpha} X^{(L)}_{\mu_1\ldots\mu_L}(k)=
\frac{L+1}{2L+1}X^{(L+1)}_{\mu_1\ldots\mu_L\alpha}(k)+
\frac{L}{2L+1}Z^{(L-1)}_{\mu_1\ldots\mu_L,\alpha}(k)\,k^2 \ ,
\ee
which is a sum rule for angular momentum $(\vec 1\oplus\vec L$
in terms of operators
$ X^{(1)}_{\alpha}(k^\perp)$ and
$X^{(L)}_{\mu_1\ldots\mu_L}(k^\perp)$.

\subsection{Projection operator}

Let us introduce a projection operator
$O^{\mu_1\ldots\mu_L}_{\nu_1\ldots \nu_L}$ for the partial wave with
the angular momentum $L$. The operator is defined by
the following relations:
\be
X^{(L)}_{\mu_1\ldots\mu_L}(k)O^{\mu_1\ldots\mu_L}_{\nu_1\ldots \nu_L}\
=\ X^{(L)}_{\nu_1\ldots \nu_L}(k)\ , \qquad
O^{\mu_1\ldots\mu_L}_{\alpha_1\ldots\alpha_L} \
O^{\alpha_1\ldots\alpha_L}_{\nu_1\ldots \nu_L}\
=\ O^{\mu_1\ldots\mu_L}_{\nu_1\ldots \nu_L}\ .
\label{proj_op}
\ee
For the sets of indices $\mu_1\ldots\mu_L$ and
$\nu_1\ldots\nu_L\;$, the operator $\hat O$ has all the properties of
the operator $\hat X^{(L)}$: it is symmetrical and traceless,
\be
O^{\mu_1\mu_2\ldots\mu_L}_{\nu_1\nu_2\ldots \nu_L}=
O^{\mu_2\mu_1\ldots\mu_L}_{\nu_1\nu_2\ldots \nu_L}=
O^{\mu_1\mu_2\ldots\mu_L}_{\nu_2\nu_1\ldots \nu_L} \  , \qquad
O^{\mu_1\mu_1\ldots\mu_L}_{\nu_1\nu_2\ldots \nu_L}=
O^{\mu_1\mu_2\ldots\mu_L}_{\nu_1\nu_1\ldots \nu_L}=0\ .
\label{17}
\ee
The projection operator $\hat O$ can be consructed as a product of
the operators \\ $X^{(L)}_{\mu_1\ldots\mu_L}(k)
X^{(L)}_{\nu_1\ldots\nu_L}(k)$ integrated over angular variables of the
momentum $k^\perp$, so we have a
convolution of the $(2L+1)$-dimensional vectors, which provide us
an irreducible representation of Lorentz group in the
$k^\perp/|k^\perp|$-space.
So,
\be
\xi(L)O^{\mu_1\ldots\mu_L}_{\nu_1\ldots \nu_L}=
\frac{1}{k^{2L}_\perp}\int\frac{d\Omega }{4\pi}\
X^{(L)}_{\mu_1\ldots\mu_L}(k)
X^{(L)}_{\nu_1\ldots\nu_L}(k)\ .
\label{2.9}
\ee
where $\xi(L)$ is a normalization factor  fixed below.
Using the definition of the projection operator
$O^{\mu_1\ldots\mu_n}_{\nu_1\ldots \nu_n}$ as well as formulae
(\ref{2.6}) and (\ref{2.8}), we obtain:
\be
k_{\mu_1}\ldots
k_{\mu_L}O^{\mu_1\ldots\mu_L}_{\nu_1\ldots \nu_L}\ =
\frac{1}{\alpha(L)}
X^{(L)}_{\nu_1\ldots\nu_L}(k)\ .
\label{2.11}
\ee
This equation represents the basic property of
the operator:
it projects any $L$-index operator into partial--wave
operator with the angular momentum $L$.

Multiplying equation
(\ref{2.9}) by the product $X^{(L)}_{\mu_1\ldots\mu_L}(q)$ and
$X^{(L)}_{\nu_1\ldots\nu_L}(q)$,
we get:
\be
\xi(L)
X^{(L)}_{\nu_1\ldots\nu_L}(q)X^{(L)}_{\nu_1\ldots\nu_L}(q)=
q^{2L}_\perp \alpha^2(L)
\int\limits_{-1}^{1}\frac{dz}{2}\; P^2_L(z)\ ,
\label{20}
\ee
that gives the normalization constant in (\ref{2.9}):
\be
\xi(L)\ =\frac{\alpha(L)}{2L+1}=
\frac{(2L-1)!!}{(2L+1)\cdot L!}\ .
\label{21}
\ee
Summation in the projection operator over upper and lower indices
performed in (\ref{2.9})
gives us the following reduction formula:
\be
O^{\mu_1\ldots\mu_{L-1}\mu_L}_{\nu_1\ldots \nu_{L-1}\mu_L}=
\frac{2L+1}{2L-1}\;O^{\mu_1\ldots\mu_{L-1}}_{\nu_1\ldots\nu_{L-1}}\ .
\label{22}
\ee
Likewise, the summation over all indices gives:
\be
O^{\mu_1\ldots\mu_L}_{\mu_1\ldots\mu_L}=2L+1 \ ,
\label{23}
\ee
that can be proven using formula (\ref{2.9}). Indeed,
$$
O^{\mu_1\ldots\mu_L}_{\mu_1\ldots\mu_L}
=\frac1{k^{2L}_\perp\xi(L)}\int\frac{d\Omega}{4\pi}X^{(L)}_{\mu_1
\ldots\mu_L}(k) X^{(L)}_{\mu_1\ldots\mu_L}(k)
=\frac{\alpha(L)}{\xi(L)}=2L+1\ .
$$
Basing on equation (\ref{2.11}), one gets
\be
X^{(L)}_{\mu_1\ldots\mu_{L-1}\mu_L}
O^{\mu_1\ldots\mu_{L-1}}_{\nu_1\ldots\nu_{L-1}}=
X^{(L)}_{\nu_1\ldots\nu_{L-1}\mu_L}\ .
\label{24}
\ee
Generally, one can write:
\be
X^{(L)}_{\mu_1\ldots\mu_{i}\mu_{i+1}\ldots\mu_L}
O^{\mu_1\ldots\mu_{i}}_{\nu_1\ldots\nu_{i}}=
X^{(L)}_{\nu_1\ldots\nu_{i}\mu_{i+1}\mu_L}\ .
\label{25}
\ee

\subsection{Four-point and five-point amplitudes for spinless
particles}

The four- and five-point amplitudes determine, correspondingly,
the scattering amplitude {\it two particles $\to$ two particles} and the
production amplitude {\it two particles $\to$ three particles}. The
latter process is considered with the impact to study  three-particle
state produced in the two-stage reaction {\it (spin-$j$) resonance
$+$ particle $\to$ three particles}.

\subsubsection{Scattering amplitude for spinless particles}

The operator
$X^{(L)}_{\mu_1\ldots\mu_L}$ is a constructive element of the
 considered here
partial--wave scattering amplitude for spinless
particles, of the type of $ SS\to SS $, or $ SP\to SP $, and so on.
We denote relative momenta of particles before and after interaction
as $q$ and $k$, correspondingly. The structure of partial-wave
amplitude with angular momentum $L=J$ is determined by the convolution of
the operators $X^{(L)}(k)$ and $X^{(L)}(q)$:
\be
A_L=a_L(s)X^{(L)}_{\mu_1\ldots\mu_L}(k)
X^{(L)}_{\mu_1\ldots\mu_L}(q)\ .
\label{ampl_scal}
\ee
Here $a_L(s)$ is the partial-wave amplitude
which depends on the total energy squared only.

The convolution
$X^{(L)}_{\mu_1\ldots\mu_L}(k)
q^\perp_{\mu_1}\ldots q^\perp_{\mu_L}$
can be written in terms of Legendre polynomials $P_L(z)$:
\be
X^{(L)}_{\mu_1\ldots\mu_L}(k)
q^\perp_{\mu_1}\ldots q^\perp_{\mu_L}
=\left(\sqrt {k^2_\perp}\sqrt{ q^2_\perp}\right)^{L} P_L(z)\  ,
\qquad
z=\frac{(k^\perp q^\perp)}{\sqrt{k^2_\perp}\sqrt{ q^2_\perp}}\  .
\label{2.6}
\ee
In the centre-of-mass system,
one has $(k^\perp q^\perp)=-(\vec k^\perp
\vec q^\perp)$ and $\sqrt{k^2_\perp} =\sqrt{-\vec k^2_\perp}
=i |\vec k_\perp |$, $\sqrt{q^2_\perp}
=i |\vec q_\perp |$.
For the low-$L$ moments, the formula (\ref{2.6}) is obvious, see Eq.
(\ref{2.2}).  For arbitrary $L$, equation (\ref{2.6}) is proven by
using (\ref{2.5}):  within a definition (\ref{2.6}),  equation
(\ref{2.5}) can be re-written as follows:
\be
P_L(z)=\frac{2L-1}{L}zP_{L-1}(z)-\frac{L-1}{L}P_{L-2}(z),
\nonumber
\ee
that is a well-known recurrent equation for the Legendre polynomials.
The convolution of the operators $X^{(L)}(k)$ and $X^{(L)}(q)$ reads:
\be
X^{(L)}_{\mu_1\ldots\mu_L}(k)X^{(L)}_{\mu_1\ldots\mu_L}(q)
=\alpha(L)
\left(\sqrt{k^2_\perp}\sqrt{ q^2_\perp}\right)^{L} P_L(z)\  ,
\label{2.8}
\ee
where $\alpha(L)$ is determined by (\ref{2.10}).
In the centre-of-mass system
$z=\cos\Theta$, and we have standard formulae for the partial-wave
expansion of the scattering amplitude.

The unitarity condition for
$a_L(s)$, which is fixed by the amplitude normalization, is determined
as follows:
\be
{\rm Im}
\; a_L(s)=\rho (s) \alpha (L)|k_\perp|^{2L}|a_L(s)|^2 + contribution\;
of\; inelastic\; channels\ .
\label{2.7}
\ee
In the right-hand side of (\ref{2.7}),  only the contribution of the
elastic channel is written explicitly, and $\rho (s)$ is the phase
space of scattered particles in this channel (recall that for the
scattering process $k^2_\perp= q^2_\perp$).

\subsubsection{Three-particle production amplitude
with resonance in the intermediate state}

The five-point amplitude for the production of three
spinless particles,
$SS\to SSS$ or $SS\to SPP$, etc., can be also represented  in terms
of operators $ X^{(L)}_{\mu_1\ldots\mu_L}$.
Let us consider here an example of the three particle production
amplitude when two final-state mesons are produced through the decay of
the resonance with spin $j$. To be definite, we consider the simplest
case of vector resonance, with $j=1$, and point out the way of
generalization for $j>1$. To consider the simplest case,
$j=1$, is in fact sufficient to trace the difference between
three-dimensional approach of Zemach \cite{Zemach} and the method
developed here.

For the two-stage reaction, of the type $SS\to VS \to PPS$,
we denote the
particle momenta in the initial and final states, correspondingly,
as $q_1,q_2$ and $k_1,k_2,k_3$; the vector resonance $(V)$ is produced
in the channel $1+2$ with the total momentum $p_{12}=k_1+k_2$
and relative momentum for the decay products $k_{12}=\frac12 (k_1-k_2)$.
Then the amplitude in the partial wave $L$ reads:
\be
a_{L,l=L+1}(s,s_{12})
X^{(L)}_{\mu_1\ldots\mu_L}(q)
X^{(L+1)}_{\mu_1\ldots\mu_L\alpha} (k^\perp_{3})
X^{(1)}_\alpha(k^\perp_{12})
\label{16a}
\ee
$$
+a_{L,l=L-1}(s,s_{12})X^{(L)}_{\mu_1\ldots\mu_L}(q)
X^{(L-1)}_{\mu_1\ldots\mu_{L-1}} (k^\perp_{3})
X^{(1)}_{\mu_{L}}(k^\perp_{12})\, ,
\nonumber
$$
where
$k^\perp_{12\mu}$ is orthogonal to to $p_{12}=k_1+k_2$,
\be
k^\perp_{12\mu}=\left
(g_{\mu\mu'}-\frac{p_{12\mu}p_{12\mu'}}{s_{12}}\right )k_{12\mu'}\ ,
\label{16b}
\ee
 $p^2_{12}=s_{12}$, and the momentum $k^\perp_3$ is orthogonal to the
total momentum $P=k_1+k_2+k_3$,
\be
k^\perp_{3\mu}=\left
(g_{\mu\mu'}-\frac{P_{\mu}P_{\mu'}}{P^2}\right )k_{3\mu'} \, .
\label{16c}
\ee
The amplitude with $l=L$ is forbidden for the processes of the
type $ SS\to PPS $ considered in (\ref{16a}). On the contrary,
when in the initial and final states, like $ SS\to VP\to SSP $,
the total particle parity changes, one has the transition
with $l=L$:
\be
a_{L,l=L}(s,s_{12}) X^{(L)}_{\mu_1\ldots\mu_L}(q)
X^{(L)}_{\mu_1\ldots\mu_{L-1}\alpha} (k^\perp_{3})
X^{(1)}_\beta(k^\perp_{12}) P_\gamma \epsilon_{\alpha\beta\gamma\mu_L}
\label{16d}
\ee
where $\epsilon_{\alpha\beta\gamma\mu_L}$ is  the totally
antisymmetrical four-tensor.

Comparing the amplitudes of Eqs. (\ref{ampl_scal}) and (\ref{16a}),
or (\ref{16d}),
one can see the common and different features in
three-dimensional approach of Zemach \cite{Zemach} and  covariant
method developed here. For the scattering amplitude (\ref{ampl_scal})
used in the centre-of-mass frame, the expressions used in both
approaches coincide. Indeed,  the four-momenta $k^\perp_\mu$ and
$q^\perp_\mu$ have space-like components only, $k^\perp_\mu=(0,\vec k)$
and $q^\perp_\mu=(0,\vec q)$, so the operators
$X^{(L)}_{\mu_1\ldots\mu_L(k)}$ and $X^{(L)}_{\mu_1\ldots\mu_L}(q)$
turn into Zemach's operators. However, for the amplitude (\ref{16a}) a
simultaneous equality to zero of operators with zero components is
impossible. In \cite{Zemach} a special procedure was suggested  for
such cases, namely, the  operator is treated in its own
centre-of-mass frame, with subsequent Lorentz boost to a needed
frame. But in the procedure developed in this paper these additional
manipulations are unnecessary.

The Lorentz boost should be also carried upon the three-particle
production amplitude considered in terms of spherical wave functions
as well as in its version suggested by \cite{Chung}.

The generalization for a higher resonance
is obvious: for the tensor resonance, $j=2$, in the reaction
$ SS\to TS \to PPS $, instead of two terms in (\ref{16a})
one has three terms  for the partial-wave amplitude:
\be
a_{L,l=L+2}(s,s_{12})
X^{(L)}_{\mu_1\ldots\mu_L}(q)
X^{(L+2)}_{\mu_1\ldots\mu_L\alpha_1\alpha_2} (k^\perp_{3})
X^{(2)}_{\alpha_1\alpha_2}(k^\perp_{12})
\label{16e}
\ee
$$
+a_{L,l=L}(s,s_{12})X^{(L)}_{\mu_1\ldots\mu_L}(q)
X^{(L)}_{\mu_1\ldots\mu_{L-1}\alpha} (k^\perp_{3})
X^{(2)}_{\alpha\mu_{L}}(k^\perp_{12})\, .
\nonumber
$$
$$
+a_{L,l=L-2}(s,s_{12})X^{(L)}_{\mu_1\ldots\mu_L}(q)
X^{(L-2)}_{\mu_1\ldots\mu_{L-2}} (k^\perp_{3})
X^{(2)}_{\mu_{L-1}\mu_{L}}(k^\perp_{12})\ ,
\nonumber
$$
In the analysis of Crysal Barrel data \cite{anis-2}, the reactions
of the proton-antiproton annihilation, of the type
$p\bar p \to resonance+P\to PPP$, are studied. Formally, the
re-writing of the above-given   formulae for the case of the $p\bar p$
initial state is simple: one needs to replace
in the amplitudes  the
 momentum operator $X^{(L)}_{\mu_1\ldots\mu_L}(q)$ by that for the
$p\bar p$ state introduced in (\ref{2.15}). A  detailed discussion of
this operator is given in the next Section.

\section{Angular-momentum operators \\ for fermion-antifermion system}

Here we consider a fermion-antifermion system,
the fermion momenta being $q_1$ and $q_2$.
The partial-wave vertex for this system with the total angular momentum
$J$, orbital momentum $L$
and total spin $S$ is determined by bilinear form
$\bar\psi(-q_2)Q^{(S,L,J)}_{\mu_1\mu_2\ldots\mu_{J-1}\mu_J}\psi(q_1)$
where $\bar\psi(-q_2)$ and $\psi(q_1)$ are bispinors
and $Q^{(S,L,J)}_{\mu_1\mu_2\ldots\mu_{J-1}\mu_J}$
is the fermion partial-wave operator. This latter operator should be
constructed with the use of the  orbital-momentum operator $X^{(L)}$ and
spin operator for fermion-antifermion system.

We have two fermion spin states, $S=0$ and $S=1$.
For the spin-0 state $J=L$, and for the spin-1 state one has
$J=L-1,L,L+1$.

For fermion operator
$Q^{(S,L,J)}_{\mu_1\mu_2\ldots\mu_{J-1}\mu_J}$, one implies
the same constraints
as for the boson one given by formulae (\ref{2.12}), (\ref{2.13}) and
(\ref{2.1}):  the fermion operator should be symmetrical,
$P$-orthogonal and traceless:
\be
Q^{(S,L,J)}_{\mu_1\mu_2\ldots\mu_{J-1}\mu_J}=
Q^{(S,L,J)}_{\mu_2\mu_1\ldots\mu_{J-1}\mu_J}\  ,\qquad
P_\mu Q^{(S,L,J)}_{\mu\mu_2\ldots\mu_{J-1}\mu_J}=0 \  , \qquad
g_{\mu_i\mu_k}Q^{(S,L,J)}_{\mu_1\ldots\mu_J}=0\ .
\label{26}
\ee
The spin-0 operator for
fermion-antifermion system, $\Gamma^{(0)}$, is proportional to the
$\gamma_5$-matrix.  We normalize $\Gamma^{(0)}$
by the condition:
\be
{\rm Sp}\  \left ( \Gamma^{(0)} (m_1+\hat q_1)
\Gamma^{(0)} (m_2-\hat q_2)\right )= 1\ ,
\label{27}
\ee
that gives
\be
\Gamma^{(0)}=\frac{i\ \gamma_5}
{\sqrt{2[s-(m_1-m_2)^2]}} \  ,
\label{28}
\ee
where $m_1$ and $m_2$ are fermion masses.
@@@
The angular--momentum operator for the spin-0 state is a product of
$\Gamma^{(0)}$ and the angular--momentum operator $X^{(J)}_
{\mu_1\ldots\mu_J}(q)$:
\be
Q^{(0,L,J)}_{\mu_1\mu_2\ldots\mu_{J-1}\mu_J}(q)=\Gamma^{(0)}
X^{(J)}_{\mu_1\ldots\mu_J}(q)  \  .
\label{29}
\ee
The spin-1
operator is constructed as follows:
\be
\Gamma_\alpha^{(1)} =
\frac{\gamma_\beta}{\sqrt{2[s-(m_1-m_2)^2]}}
\left(g^\perp_{\beta\alpha}-\frac{4s\;q^\perp_\beta
q^\perp_\alpha}{[(m_1+m_2)^2+(m_1+m_2)\sqrt s
\ ][s-(m_1-m_2)^2]}\right) .
\label{30}
\ee
For equal masses, $m_1=m_2=m$, the spin-1
operator reads:
\be
\Gamma_\alpha^{(1)} =
\frac{1}{\sqrt{2s}}
\left(\gamma^\perp_{\alpha} -
\frac{\hat q q_\alpha }{m(m+\frac{\sqrt{s}}{2})}
\right)=
\frac{1}{\sqrt{2s}}
\left(\gamma^\perp_{\alpha} -
\frac{ q_\alpha }{m+\frac{\sqrt{s}}{2}}\right)\ .
\label{30a}
\ee
Here we have used that
\be
\hat q=\frac {\hat k_1 -\hat k_2}{2}=m\ .
\label{30b}
\ee
The
operator $\Gamma^{(1)}_\alpha$ is orthogonal to the total momentum
$P$,
\be
P_\alpha
\Gamma_\alpha^{(1)} =0  \ ,
\label{31}
\ee
and normalized by the condition:
\be
{\rm Sp}\;
\left (\Gamma_\alpha^{(1)} (m_1+\hat q_1) \Gamma_\beta^{(1)} (m_2-\hat
q_2)\right )= g^\perp_{\alpha\beta} \ .
\label{normg}
\ee
The operators $\Gamma_\alpha^{(1)}$ and $\Gamma^{(0)}$ are orthogonal
to each other in the spin space:
\be
{\rm Sp}\;
\left (\Gamma_\alpha^{(1)} (m_1+\hat q_1) \Gamma^{(0)} (m_2-\hat
q_2)\right )= 0 \ .
\label{33}
\ee
It means that, if initial fermions are not polarized, the spin-0
and spin-1 states do not interfere with each other in the differential
cross section.

The spin-1 state with $J=L-1$ is constructed from
the operator $\Gamma_\alpha^{(1)}$ and angular--momentum operator
$X^{(J+1)}_{\mu_1\ldots\mu_J\mu_{J+1}}$. It carries $J$ indices, so
two indices, one from $\Gamma_\alpha^{(1)}$
and another from $X^{(J+1)}_{\mu_1\ldots\mu_J\mu_{J+1}}$,
are to be absorbed, that can be done with the help of the metric tensor
$g_{\alpha\mu_{J+1}}$:
\be
Q^{(1,L,J=L-1)}_{\mu_1\ldots\mu_J}(q)=\Gamma_\alpha^{(1)}
X^{(L)}_{\mu_1\ldots\mu_{L-1}\alpha}(q) \  .
\label{34}
\ee
For $J=L$,
the construction of the operator $Q^{(1,L,J=L)}_{\mu_1\ldots\mu_J}$
is performed by using
antisymmetrical tensor $\varepsilon_{\mu\nu_1\nu_2\nu_3}$:
the operator
$\Gamma_\nu X^{(J)}_{\mu_1\mu_2\ldots\mu_J}$
must have the same number of indices as the angular--momentum operator.
The only possible non-zero combination of antisymmetrical
tensor, operators $X^{(J)}$ and $\Gamma^{(1)}$
is given by the following convolution:
\be
\varepsilon_{\mu_1\nu_1\nu_2\nu_3}P_{\nu_1}
\Gamma_{\nu_2}^{(1)}
X^{(J)}_{\nu_3\mu_2\ldots\mu_J}(q)  \ .
\label{2.14}
\ee
However, the operator entering equation (\ref{2.14}), being
$P$-orthogonal and traceless, is not symmetrical. The symmetrization
can be performed by using the tensor $Z^{(J-1)}_{\nu\mu_2\ldots\mu_J
,\alpha}$ instead of $X^{(J)}_{\nu\mu_2\ldots\mu_J}$. In this way, we
get:
\be
Q^{(1,L,J=L)}_{\mu_1\ldots\mu_J}(q)=
\varepsilon_{\alpha\nu_1\nu_2\nu_3}P_{\nu_1} \Gamma_{\nu_2}^{(1)}
Z^{(J)}_{\nu_3\mu_1\ldots\mu_J,\alpha} (q)\ .
\label{36}
\ee
Following the same procedure, we can easily construct the operator for
the total angular momentum $J=L+1$. One has:
\be
Q^{(1,L,J=L+1)}_{\mu_1\ldots\mu_J}(q)=
\Gamma_\alpha^{(1)}
Z^{(J-1)}_{\mu_1\ldots\mu_J,\alpha}(q)\ .
\label{37}
\ee

\section{Amplitude for the fermion-antifermion annihilation
into two spinless mesons}

The partial--wave amplitude
for the $p\bar p $-annihilation
into scalar $(S)$ and pseudoscalar $(P)$ mesons
($p\bar p \to SS, \  PP, \; SP$) is given in a general form by formula
(\ref{2.15}).  To be illustrative we present in this section the
formulae which are helpful for the analysis of these reactions.

For the fermion-antifermion state with the total zero spin,
 $S=0$, the amplitude is determined by the following angle-dependent
factor:
\be
\Gamma^{(0)}
X^{(J)}_{\mu_1\ldots\mu_J}(q) X^{(J)}_{\mu_1\ldots\mu_J}(k)= \Gamma^{(0)}
\alpha(J) \bigg(\sqrt{q^2_\perp}\bigg)^J \bigg(\sqrt{k^2_\perp}\bigg)^J
P_J(z)\ .
\ee
 Recall that $q$ is the relative momentum of  fermions
and $k$ is that of bosons and $z=(k^\perp
q^\perp)/(\sqrt{k^2_\perp}\sqrt{q^2_\perp})$.

In the spin-1 case, when $L=J+1$,
we have the following convolution:
\be
X^{(J)}_{\mu_1\ldots\mu_J}(k)X^{(J+1)}_{\mu_1\ldots\mu_J\beta}(q)
=\alpha(J)\left(\sqrt{k^2_\perp}\right)^J
\left(\sqrt{q^2_\perp}\right)^{J+1}
\left[ A_J\frac{k_\beta^\perp}{\sqrt{k^2_\perp}}
+B_J\frac{q^\perp_\beta}{\sqrt{q^2_\perp}}      \right]\ ,
\label{a+b}
\ee
$A_J$ and $B_J$ being the coefficients which depend on
$z$ only.
They can be found by multiplying
both sides of equation (\ref{a+b}) by $q^\perp_\beta$ and
$k^\perp_\beta$:
$$
\alpha(J)\left(\sqrt{k^2_\perp}\right)^J\left(\sqrt{q^2_\perp}
\right)^J\;q^2_\perp P_{J}(z)=\alpha(J)\left(\sqrt{k^2_\perp}
\right)^J\left(\sqrt{q^2_\perp}\right)^{J+1} \left[A_J
\frac{k^\perp q^\perp}{\sqrt{k^2_\perp}} +B_J
\sqrt{q^2_\perp}\right]\ ,
%\label{eqab1}
$$
$$
\alpha(J)\left(\sqrt{k^2_\perp}\right)^{J+1}\left(\sqrt{q^2_\perp}
\right)^{J+1}\; P_{J+1}(z)=\alpha(J)\left(
\sqrt{k^2_\perp}\right)^J\left(\sqrt{q^2_\perp}\right)^{J+1}
\left[A_J\sqrt{k^2_\perp}+B_J
\frac{k^\perp q^\perp}{\sqrt{q^2_\perp}}   \right]\ .
%\label{eqab2}
$$
One easily gets for $A_J$ and $B_J$:
\be
A_J =\ \frac{P_{J+1}(z)-z\,P_J(z)}{1-z^2}\ , \qquad B_J=\
\frac{P_J(z)-zP_{J+1}(z)}{1-z^2} \  .
\label{ab_fin}
\ee
For the first partial waves, one has:
\be
 A_1=-\frac12,&A_2=-z ,  &A_3=-\frac38(5z^2-1),
\nonumber \\
B_1=\frac32\ z\ ,&B_2=\frac12(5z^2-1)\ , &B_3=\frac58(7z^2-3z)\ .
\ee
For $L=J-1$, the angle-dependent factor contains the
convolution of
$Z^{(J-1)}_{\mu_1\ldots\mu_J,\alpha}(q)$ and
$X^{(J)}_{\mu_1\ldots\mu_J}(k)$. It reads:
\be
&&Z^{(J-1)}_{\mu_1\ldots\mu_J,\alpha}(q)X^{(J)}_{\mu_1\ldots\mu_J}(k)
=\frac{2J-1}{J^2}\biggl( \sum^J_{i=1}g^\perp_{\mu_i\alpha}
X^{(J-1)}_{\mu_1\ldots\mu_{i-1}\mu_{i+1}\ldots\mu_J}(q)-
\nonumber \\
&&-\frac2{2J-1}\sum^J_{i,j=1 \atop i<j}
g^\perp_{\mu_i\mu_j}X^{(J-1)}_
{\mu_1\ldots\mu_{i-1}\mu_{i+1}\ldots\mu_{j-1}\mu_{j+1}
\ldots\mu_J\alpha}(q)\biggr) X^{(J)}_{\mu_1\ldots\mu_J}(k)
 \nonumber \\
&&=\alpha(J)\left(\sqrt{k^2_\perp}\right)^{J}
\left(\sqrt{q^2_\perp}\right)^{J-1}\left[
A_{J-1}\frac{q^\perp_\alpha}{\sqrt{q^2_\perp}}
+B_{J-1}\frac{k^\perp_\alpha}{\sqrt{k^2_\perp}}
\right] \ .
\label{z*x}
\ee
The angular structure of the amplitude
{\it fermion+antifermion $\to$ two mesons }
when $L=J\pm 1$ is determined by the following factors:
\be
&&(^3(J+1)_J\to
J)=\frac{1}{\alpha(J)}\frac{J+1}{2J+1} \Gamma_\alpha
X^{(J+1)}_{\mu_1\ldots\mu_J\alpha}(q) X^{(J)}_{\mu_1\ldots\mu_J}(k)\ ,
\nonumber \\
&&(^3(J-1)_J\to J)=\frac{1}{\alpha(J)}\frac{J}{2J+1}
\Gamma_\alpha Z^{(J-1)}_{\mu_1\ldots\mu_J,\alpha}(q)
X^{(J)}_{\mu_1\ldots\mu_J}(k)\ ,
\label{mat_el_n}
\ee
where
\be
&&\Gamma_\beta X^{(J+1)}_{\mu_1\ldots\mu_J\beta}(q)
X^{(J)}_{\mu_1\ldots\mu_J}(k)=
\Gamma_\beta \left(\sqrt{q^2_\perp}\right)^{J+1}\left(
\sqrt{k^2_\perp}\right)^J\alpha(J)\left[
A_J\frac{k^\perp_\beta}{\sqrt{k^2_\perp}}+
B_J\frac{q^\perp_\beta}{\sqrt{q^2_\perp}}
\right]\ ,
 \\
&&\Gamma_\beta Z^{(J-1)}_{\mu_1\ldots\mu_J,\beta}(q)
X^{(J)}_{\mu_1\ldots\mu_J}(k)=
\Gamma_\beta
\left(\sqrt{q^2_\perp}\right)^{J-1}\left(
\sqrt{k^2_\perp}\right)^J\alpha(J)\left[
A_{J-1}\frac{q^\perp_\beta}{\sqrt{q^2_\perp}}+
B_{J-1}\frac{k^\perp_\beta}{\sqrt{k^2_\perp}}
\right]\ .
\nonumber
\label{mat_el}
\ee
The normalization of  angular--momentum operators is a matter of
convenience.  To this aim we
average the matrix element squared over fermion polarizations and
angles of the boson system. We have
\be
&&\langle (^3(J+1)_J\to J)^2\rangle=-\frac12
(q^2_\perp)^{J+1}(k^2_\perp)^J\ \frac{J+1}{(2J+1)^2} \ ,
\nonumber \\
&&\langle(^3(J-1)_J\to J)^2\rangle=-\frac12
(q^2_\perp)^{J-1} (k^2_\perp)^J\ \frac{J}{(2J+1)^2}\ .
\ee
The normalization of matrix elements given by (\ref{mat_el_n})
can be easily checked
in the centre-of-mass system, where the angle  $z$ has
a simple definition (recall that in this system
${\bf k}^2=-k^2_\perp$, ${\bf q}^2=-q^2_\perp$).

The amplitudes for the transition
from fermion states with $L=J\pm 1$ to the boson state $J$ via
Breit-Wigner resonance are equal to:
\be
&A(^3(J+1)_J\to J)&=
\frac{J+1}{2J+1}\; \sqrt{q^2_\perp}
\Gamma_\alpha\left[
A_J\frac{k^\perp_\alpha}{
\sqrt{k^2_\perp}}+B_J\frac{q^\perp_\alpha}{\sqrt{q_\perp^2}}
\right]\; BW(J+1,J)\ ,
\\
&A(^3(J-1)_J\to J)&=
\frac{J}{2J+1}\; \frac{1}{\sqrt{q^2_\perp}}
\Gamma_\alpha\left[A_{J-1}\frac{q^\perp_
\alpha}{\sqrt{q^2_\perp}}+B_{J-1}\frac{k^\perp_\alpha}{
\sqrt{k^2_\perp}}\right]\; BW(J-1,J)\ .
\label{ampl_fs}
\nonumber
\ee
Here the kinematical factors $q^2_\perp$, $k^2_\perp$
are included into the definition of the Breit-Wigner amplitude
$BW$
as well as coupling constants of the resonance decaying into
two-fermion and two-boson channels:
\be
&&BW(J+1,J)=\frac{(\sqrt{q^2_\perp})^{J} (\sqrt{k^2_\perp})^{J}}
{F_{J+1}(q^2_\perp)F_J(k^2_\perp)}\; \frac{C_{J+1}C}{M^2-s-i\rho
M\Gamma}\ , \nonumber \\
&&BW(J-1,J)=\frac{(\sqrt{q^2_\perp})^{J}
(\sqrt{k^2_\perp})^{J}}
{F_{J-1}(q^2_\perp)F_J(k^2_\perp)}\;
\frac{C_{J-1}C}{M^2-s-i\rho M\Gamma}\ .
\ee
Here $F$ are the form factors, $C$ is the coupling constant of   the
resonance to boson state, and $C_{J+1}$ and $C_{J-1}$ are the couplings
to fermion states, with $L=J+1$ and $L=J-1$.

The complete amplitude for the transition $ fermion+antifermion
\to boson \;
states$ is equal to the sum of partial--wave amplitudes:
\be
A_{tot}= \sqrt{q^2_\perp}
\Gamma_\alpha\left[ \tilde
V_1\frac{q^\perp_\alpha}{\sqrt{q^2_\perp}}+ \tilde
V_2\frac{k^\perp_\alpha}{\sqrt{k_\perp^2}} \right]\ ,
\label{v1v2}
\ee
where:
\be
\tilde V_1=\sum^n_{J=0}\frac{1}{2J+1}
\bigg[J\;A_{J-1}BW(J-1,J)\frac{1}{q^2_\perp}+B_{J}(J+1)BW(J+1,J) \bigg
]\ , \nonumber \\
\tilde V_2=\sum^n_{J=1}\frac{1}{2J+1}
\bigg[J\;B_{J-1}BW(J-1,J)\frac{1}{q^2_\perp}+A_{J}(J+1)BW(J+1,J) \bigg
]\ .
\label{v1v2_d}
\ee

Generalization of the formulae (\ref{ampl_fs})--(\ref{v1v2_d}) for the
case, when partial wave amplitude describes several resonances
and cerain background, is
obvious; one should substitute:
\be
BW \rightarrow \sum_a BW_a+background \ .
\label{backgr}
\ee

Consider the case when one fermion with fixed momentum interacts
with another fermion  producing by annihilation the boson states
(e.g. $\bar p p\to \pi^+\pi^-$, $\pi^0\pi^0$, $\eta\eta$, and so on).
If  fermions are not polarized, the cross section of
such a process is equal to the amplitude squared multiplied by the phase
volume of final particles.  We have for the amplitude squared:
\be
|A_{tot}|^2=\frac{-q^2_\perp}{2}\bigg [|\tilde V_1|^2+|\tilde V_2|^2+
z(\tilde V_1\tilde V_2^*+\tilde V_1^*\tilde V_2) \bigg ]\ .
\label{A2}
\ee
Sometimes it is convenient to use the other functions:
\be
V_1=\tilde V_1
+z\tilde V_2, \qquad \qquad V_2=\sqrt{1-z^2}\tilde V_2\ .
\label{v_fin}
\ee
Here the amplitude squared has  rather simple form:
\be
|A_{tot}|^2=\frac{-q^2_\perp}{2} \bigg [ |V_1|^2+|V_2|^2 \bigg ]\ .
\label{A2fin}
\ee
In the next section we will show that the amplitudes $V_1$ and $V_2$ are
connected with the helicity amplitudes.

\subsection*{The polarized target fermion}

Consider the case, when one of  initial fermions is polarized (the
reaction $pp \to \pi^+\pi^-$, with polarized target proton, was studied
in \cite{Hasan}).
The fermion loop diagram for this case is equal to:
\be
N&=&-\frac12
{\rm Sp}\left[\Gamma_\mu(m-\hat q_1)(1-\gamma_5\hat S)
\Gamma_\nu(m+q_2) \right]\nonumber \\
&=&-\frac12 \Delta_{\mu\alpha}\Delta_{\nu\beta}
Sp\left[\gamma_\mu(m-\hat q_1)(1-\gamma_5\hat S)
\gamma_\nu(m+q_2)\right ]\ .
\ee
Here $\hat S = S_\mu \gamma_\mu$ where $S_\mu$ is the polarization
vector. After simple calculations we obtain:
\be
N=-\frac12 g^\perp_{\mu\nu} - i
\frac{m S_\alpha P_\xi}{s}
\left [ \varepsilon_{\alpha\mu\nu\xi}+
\frac{2q_\beta}{m(2m+\sqrt s)}    
(q_\nu\varepsilon_{\alpha\beta\mu\xi}-
q_\mu\varepsilon_{\alpha\beta\nu\xi})\right ].
\ee
The amplitude squared is equal to:
\be
|A_{tot}|^2&=&-\frac{q^2}{2}\left [
\tilde V_1^2+\tilde V_2^2+
z(\tilde V_1 \tilde V_2^*-\tilde V_1^*\tilde V_2)+\right .
\nonumber \\
&+&\left .
i\frac{m \varepsilon_{\alpha\mu\nu\xi}S_\alpha q_\mu k_\beta
P_\xi} {s\sqrt{q_\perp^2}\sqrt{k_\perp^2}}\,(\tilde V_1 \tilde
V_2^*-\tilde V_1^*\tilde V_2)(1-Xq_\perp^2) \right ]\ .
\ee
After simple calculations we get the final expression
in terms of the $V_i$ amplitudes:
\be
|A_{tot}|^2=|A_{tot}^{unpolirized}|^2-
\frac{q_\perp^2}{\sqrt s}
\frac{\varepsilon_{\alpha\mu\nu\xi}S_\alpha q_\mu k_\nu P_\xi}
{\sqrt{q^2_\perp}\sqrt{k^2_\perp}}\,{\rm Im}(\tilde V_1\tilde V_2^*)
\ee
As an example, let us calculate the case when polarization vector
is directed along the $y$-axis:
\be
S=(0;0,1,0)\ , \qquad P=(\sqrt s ;0,0,0)\ ,\qquad q=(q_0;0,0,|\vec q|)\
.
\ee
Then
\be
\varepsilon_{\alpha\mu\nu\xi}S_\alpha q_\mu k_\nu P_\xi=
-|\vec q||\vec k|\sqrt {1-z^2} \cos \Phi\ ,
\ee
where $\Phi$ is the angle between the $x$-axis and the projection of
${\bf k}$ on the $x,y$-plane.
Then we have the following expression for the amplitude squared:
\be
|A_{tot}|^2=-\frac{q^2_\perp}{2}\left [ V_1^2+V_2^2+2\,{\rm Im}
(V_1V^*_2)\cos \Phi\right ]\ .
\ee

\section{Operators for the two-photon states}

Similarly to the two-meson and fermion-antifermion states, let us
denote the moment operator for the two--photon state as
\be
G^{(J)}_{\mu_1\mu_2 \; .... \mu_J} (q_1,q_2)\ ,
\ee
where $q_1$ and $q_2$ are the photon momenta and $q^2_1=q^2_2=0$. For
the reaction $\gamma(q_1)\gamma(q_2)\to hadrons $ this operator should
be convoluted with relevant hadronic moment operator.

There is a specific feature in the construction of moment operator for
the $\gamma(q_1)\gamma(q_2)$ state that is a consequence of gauge
invariance: the photon is polarized in the plane orthogonal to the
collision plane of photons. For example, if photons collide along the
$z$-axis, the photon polarizations lie in the plane $(x,y)$. This
requirement reduces the set of allowed states; that is clearly seen by
using the state $J=0$ as an example.

The state $J=0$ is constructed as follows:
\be
G^{(0)}(q_1,q_2)=
\epsilon^{(1)}_{\alpha}S^{(0)}_{\alpha\beta}\epsilon^{(2)}_{\beta}
\equiv S^{(0)} \ ,
\ee
where $\epsilon^{(i)}$ is the photon polarization, with
$(\epsilon^{(i)}q_i)=0$. The moment operator $S^{(0)}_{\alpha\beta}$
has the following structure:
\be
S^{(0)}_{\alpha\beta}\sim
g^{\perp}_{\alpha\beta}+B
(q_\alpha q_\beta-\frac13
g^{\perp}_{\alpha\beta}q^2)\ .
\label{star}
\ee
The first term presents $(S=0,L=0)$ state and the second one
gives the
$(S=2,L=2)$ state, while $B$ is
a constant which should be found from
gauge requirement.
The gauge constraint tells: after convoluting (\ref{star})
with $q_{1\alpha}$ (or  $q_{2\beta}$), it should be equal zero:
\be
q^{\perp}_{1\beta}+B\left [(q_1q)q_\beta-\frac13
q^{\perp}_{1\beta}q^2\right ]=0\ .
\label{twostar}
\ee
As a result we find $B=-((q_1q)-1/3 \; q^2)^{-1}=-3/(2q^2)$, and
equation  (\ref{star}) gives $S^{(0)}_{\alpha\beta}\sim
g^{\perp\perp}_{\alpha\beta}$
where
\be
g^{\perp\perp}_{\alpha\beta}=g_{\alpha\beta}-
(q_{1\alpha} q_{2\beta}+q_{2\alpha} q_{1\beta})\frac{1}{(q_1q_2)}\ .
\ee
We define
\be
S^{(0)}_{\alpha\beta}=
g^{\perp\perp}_{\alpha\beta}\ ,
\ee
Sometimes it is more convenient to rewrite
$g^{\perp\perp}_{\alpha\beta}$ in the terms of the photon
total and relative momenta, $P=q_1+q_2$ and $q$:
\be
g^{\perp\perp}_{\alpha\beta}=g_{\alpha\beta}-
\frac{q_\alpha q_\beta}{q^2} -
\frac{P_\alpha P_\beta}{P^2}
\label{gpp}
\ee
with
\be
q_\mu=\frac12 (q_1-q_2)_\nu g^\perp_{\mu\nu},
\qquad g^\perp_{\mu\nu} =g_{\mu\nu}- \frac{P_\mu P_\nu}{P^2} \; .
\ee
We see that the operator
$ S^{(0)}_{\alpha\beta}$ contains
$D$-wave contribution, and it is a direct result of the gauge
constraint.  However the $D$-wave contribution turns into zero in
$G^{(0)}(q_1,q_2)=S^{(0)}$, so the operator $S^{(0)}_{\alpha\beta}$ can be
conventionally named the spin-$0$ operator.

Another operator, $S^{(2)}_{\alpha_1 \alpha_2\; ,\; \mu_1\mu_2}$,
we construct by using the
orthogonality requirement with respect to the spin-$0$ operator:
\be
S^{(0)}_{\alpha_1 \alpha_2}
S^{(2)}_{\alpha_1 \alpha_2\; ,\; \mu_1\mu_2} =0\; .
\ee
The operator $S^{(2)}_{\alpha_1 \alpha_2\; ,\; \mu_1\mu_2}$
can be named, also conventionally, the spin-$2$
operator. We write:
\be
 S^{(2)}_{\alpha_1 \alpha_2\; ,\; \mu_1\mu_2}
=
g^{\perp\perp}_{\mu_1\alpha_1} g^{\perp\perp}_{\mu_2\alpha_2} +
g^{\perp\perp}_{\mu_1\alpha_2} g^{\perp\perp}_{\mu_2\alpha_1}
-g^{\perp\perp}_{\mu_1\mu_2}
g^{\perp\perp}_{\alpha_1\alpha_2} \ .
\ee
The state with $J=2$, which corresponds to this operator, reads:
\be
G^{(2)}_{\mu_1\mu_2}(q_1,q_2) =
\epsilon^{(1)}_{\alpha_1}
\epsilon^{(2)}_{\alpha_2}
 S^{(2)}_{\alpha_1 \alpha_2\; ,\; \mu_1\mu_2}\equiv
S^{(2)}_{\mu_1\mu_2} \ .
\ee
All the remaining even-state operators we construct from the operators
$ S^{(0)}_{\alpha\beta}$ and $S^{(2)}_{\alpha_1 \alpha_2\; ,\;
\mu_1\mu_2}$.

The tensor $g^{\perp\perp}_{\alpha\beta}$ determines
the completeness condition for the photon
polarization vectors used for the cross
section calculations.
The completeness condition reads as
\be \sum_i
\epsilon^{(a)}_{\alpha}(i)\epsilon^{(a)}_{\beta}(i)=
g^{\perp\perp}_{\alpha\beta}\; .
\ee

\subsection{Even-state operators, $P=+$}

Now let us construct the operators of the even states with an arbitrary
$J$. The orbital  momentum of the photons should be also even:
$L=0,2,4,...$ :
in this way we have the $P$-even  states, $P=+$.

{\bf (i) Operators $S^{(0)}X^{(L)}_{\mu_1...\mu_L}(q)$ }

There are the following initial state operators which are
built with use of $S^{(0)}$:
\be
S^{(0)}X^{(L)}_{\mu_1...\mu_L}(q)\; ,
\ee
or, being more definite,
\be
&0^{++}:&\qquad S^{(0)}
\nonumber \\
&2^{++}:&\qquad S^{(0)}X^{(2)}_{\mu\nu} (q)
\nonumber \\
&4^{++}:&\qquad S^{(0)}X^{(4)}_{\mu\nu\eta\xi} (q)
\ee

{\bf (ii) Operators
$S^{(2)}_{\mu_1\mu_2}X^{(L)}_{\mu_3...\mu_{L+2}}(q)$ }

For construction operators with
$S^{(2)}_{\mu_1\mu_2}$ we use the product
\be
S^{(2)}_{\mu_1\mu_2}X^{(L)}_{\mu_3...\mu_{L+2}}(q)\ .
\label{op_s02}
\ee
and expand it in a set of states with $J=L-2,\; L-1,\; L,\; L+1,\;
L+2$.
The simplest operator is that for $L=0$:
\be
2^{++}:&\qquad
S^{(2)}_{\mu_1\mu_2}\ .
\label{L=0}
\ee
To clarify the situation
with $L\ne 0$, let us consider the case $S=2, \; L=2$ in detail:
\be
S^{(2)}_{\mu_1\mu_2}X^{(2)}_{\mu_3\mu_4}(q)\ .
\ee
The $0^{++}$ state operator, being constucted as
$ S^{(2)}_{\mu_1\mu_2}X^{(2)}_{\mu_1\mu_2}(q)$ is equal to zero:
$$
\left ( \epsilon^{(1)}_{\perp\mu_1}
\epsilon^{(2)}_{\perp\mu_2} +
\epsilon^{(1)}_{\perp\mu_2}
\epsilon^{(2)}_{\perp\mu_1}-g^{\perp\perp}_{\mu_1\mu_2}
(\epsilon^{(1)}_\perp\epsilon^{(2)}_\perp)\right )
\frac{3}{2}\left (
q_{\mu_1} q_{\mu_2}-\frac13
g^{\perp}_{\mu_1\mu_2}q^2 \right ) =0.
$$
The $1^{++}$ state does not appears  for the transverse polarized
photons as well:
$$
\varepsilon_{\mu\alpha_1\alpha_2\alpha_3} P_{\alpha_1}
X^{(2)}_{\alpha_2 \beta}(q)S^{(2)}_{\alpha_3\beta}
=
\frac{3}{4}\left(\varepsilon_{\mu\alpha_1\alpha_2\alpha_3} P_{\alpha_1}q_{\alpha_2}
\epsilon^{(1)}_{\perp\alpha_3} (q_1 \epsilon^{(2)}_\perp)-
\varepsilon_{\mu\alpha_1\alpha_2\alpha_3} P_{\alpha_1}q_{\alpha_2}
\epsilon^{(2)}_{\perp\alpha_3} (q_2 \epsilon^{(1)}_\perp)\right)=0\ .
$$
The $2^{++}$ state operator, being constucted as
$S^{(2)}_{\mu_1\mu}X^{(2)}_{\mu\mu_2}(q)$, gives $S^{(2)}_{\mu_1\mu_2}$,
up to a numerical factor:
$$
S^{(2)}_{\mu_1\mu}X^{(2)}_{\mu\mu_2}(q)=
\left ( \epsilon^{(1)}_{\perp\mu_1}
\epsilon^{(2)}_{\perp\mu} +
\epsilon^{(1)}_{\perp\mu}
\epsilon^{(2)}_{\perp\mu_1}-g^{\perp\perp}_{\mu_1\mu}
(\epsilon^{(1)}_\perp\epsilon^{(2)}_\perp)\right )
\left (
q_{\mu} q_{\mu_2}-\frac13
g^{\perp}_{\mu\mu_2}q^2 \right ) =
$$
$$
-\frac{q^2}{2}
\left ( \epsilon^{(1)}_{\perp\mu_1}
\epsilon^{(2)}_{\perp\mu_2} +
\epsilon^{(1)}_{\perp\mu_2}
\epsilon^{(2)}_{\perp\mu_1}-g^{\perp\perp}_{\mu_1\mu_2}
(\epsilon^{(1)}_\perp\epsilon^{(2)}_\perp)\right )\sim
S^{(2)}_{\mu_1\mu_2} \ .
$$
So we do not need to use this convolution, it is taken into account
in (\ref{L=0}) as well.

The $3^{++}$ state operator is constructed as
\be
&3^{++}:& \qquad
\varepsilon_{\mu_1\alpha_1\alpha_2\alpha_3} P_{\alpha_1}
X^{(2)}_{\alpha_2 \mu_2}(q)S^{(2)}_{\alpha_3\mu_3}\ ,
\nonumber
\label{3++}
\ee
with subsequent symmetrization over indices $\mu_1 ,\mu_2 ,\mu_3$
and making the operator traceless. One has for this operator:
\be
3^{++}:\;\frac{3}{2}\left[\epsilon^{(1)}_{\perp\mu_1}q_{\mu_2}
\varepsilon_{\mu_3\alpha_1\alpha_2\alpha_3} P_{\alpha_1}
q_{\alpha_2} \epsilon^{(2)}_{\perp\alpha_3}
+\epsilon^{(2)}_{\perp\mu_1}q_{\mu_2}
\varepsilon_{\mu_3\alpha_1\alpha_2\alpha_3} P_{\alpha_1}
q_{\alpha_2} \epsilon^{(1)}_{\perp\alpha_3}
\right .
\\
\nonumber
                +\epsilon^{(1)}_{\perp\mu_1}q_{\mu_3}
\varepsilon_{\mu_2\alpha_1\alpha_2\alpha_3} P_{\alpha_1}
q_{\alpha_2} \epsilon^{(2)}_{\perp\alpha_3}
+
      \epsilon^{(2)}_{\perp\mu_1}q_{\mu_3}
\varepsilon_{\mu_2\alpha_1\alpha_2\alpha_3} P_{\alpha_1}
q_{\alpha_2} \epsilon^{(1)}_{\perp\alpha_3}
\\
\nonumber
                 +\epsilon^{(1)}_{\perp\mu_2}q_{\mu_1}
\varepsilon_{\mu_3\alpha_1\alpha_2\alpha_3} P_{\alpha_1}
q_{\alpha_2} \epsilon^{(2)}_{\perp\alpha_3}
+
      \epsilon^{(2)}_{\perp\mu_2}q_{\mu_1}
\varepsilon_{\mu_3\alpha_1\alpha_2\alpha_3} P_{\alpha_1}
q_{\alpha_2} \epsilon^{(1)}_{\perp\alpha_3}
\\
\nonumber
                 +\epsilon^{(1)}_{\perp\mu_2}q_{\mu_3}
\varepsilon_{\mu_1\alpha_1\alpha_2\alpha_3} P_{\alpha_1}
q_{\alpha_2} \epsilon^{(2)}_{\perp\alpha_3}
+
      \epsilon^{(2)}_{\perp\mu_2}q_{\mu_3}
\varepsilon_{\mu_1\alpha_1\alpha_2\alpha_3} P_{\alpha_1}
q_{\alpha_2} \epsilon^{(1)}_{\perp\alpha_3}
\\
\nonumber
                  +\epsilon^{(1)}_{\perp\mu_3}q_{\mu_1}
\varepsilon_{\mu_2\alpha_1\alpha_2\alpha_3} P_{\alpha_1}
q_{\alpha_2} \epsilon^{(2)}_{\perp\alpha_3}
+
      \epsilon^{(2)}_{\perp\mu_3}q_{\mu_1}
\varepsilon_{\mu_2\alpha_1\alpha_2\alpha_3} P_{\alpha_1}
q_{\alpha_2} \epsilon^{(1)}_{\perp\alpha_3}
\\
\nonumber
                  +\epsilon^{(1)}_{\perp\mu_3}q_{\mu_2}
\varepsilon_{\mu_1\alpha_1\alpha_2\alpha_3} P_{\alpha_1}
q_{\alpha_2} \epsilon^{(2)}_{\perp\alpha_3}
+
      \epsilon^{(2)}_{\perp\mu_3}q_{\mu_2}
\varepsilon_{\mu_1\alpha_1\alpha_2\alpha_3} P_{\alpha_1}
q_{\alpha_2} \epsilon^{(1)}_{\perp\alpha_3}
\\
\nonumber
-\frac23\left(
g^\perp_{\mu_1\mu_2}\left(
\epsilon^{(1)}_{\perp\lam}q_{\mu_3}
\varepsilon_{\lam\alpha_1\alpha_2\alpha_3} P_{\alpha_1}
q_{\alpha_2} \epsilon^{(2)}_{\perp\alpha_3}
+
\epsilon^{(2)}_{\perp\lam}q_{\mu_3\lam}
\varepsilon_{\lam\alpha_1\alpha_2\alpha_3} P_{\alpha_1}
q_{\alpha_2} \epsilon^{(1)}_{\perp\alpha_3}
\right .\right .
\\
\nonumber
\left .
+\epsilon^{(1)}_{\perp\mu_3}q_{\lam}
\varepsilon_{\lam\alpha_1\alpha_2\alpha_3} P_{\alpha_1}
q_{\alpha_2} \epsilon^{(2)}_{\perp\alpha_3}
+
\epsilon^{(2)}_{\perp\mu_3}q_{\lam}
\varepsilon_{\lam\alpha_1\alpha_2\alpha_3} P_{\alpha_1}
q_{\alpha_2} \epsilon^{(1)}_{\perp\alpha_3}\right)
\\
\nonumber
+g^\perp_{\mu_1\mu_3}\left(
\epsilon^{(1)}_{\perp\lam}q_{\mu_2}
\varepsilon_{\lam\alpha_1\alpha_2\alpha_3} P_{\alpha_1}
q_{\alpha_2} \epsilon^{(2)}_{\perp\alpha_3}
+
\epsilon^{(2)}_{\perp\lam}q_{\mu_2\lam}
\varepsilon_{\lam\alpha_1\alpha_2\alpha_3} P_{\alpha_1}
q_{\alpha_2} \epsilon^{(1)}_{\perp\alpha_3}
\right .
\\
\nonumber
\left .
+\epsilon^{(1)}_{\perp\mu_2}q_{\lam}
\varepsilon_{\lam\alpha_1\alpha_2\alpha_3} P_{\alpha_1}
q_{\alpha_2} \epsilon^{(2)}_{\perp\alpha_3}
+
\epsilon^{(2)}_{\perp\mu_2}q_{\lam}
\varepsilon_{\lam\alpha_1\alpha_2\alpha_3} P_{\alpha_1}
q_{\alpha_2} \epsilon^{(1)}_{\perp\alpha_3}\right)
\\
\nonumber
+g^\perp_{\mu_2\mu_3}\left(
\epsilon^{(1)}_{\perp\lam}q_{\mu_1}
\varepsilon_{\lam\alpha_1\alpha_2\alpha_3} P_{\alpha_1}
q_{\alpha_2} \epsilon^{(2)}_{\perp\alpha_3}
+
\epsilon^{(2)}_{\perp\lam}q_{\mu_1\lam}
\varepsilon_{\lam\alpha_1\alpha_2\alpha_3} P_{\alpha_1}
q_{\alpha_2} \epsilon^{(1)}_{\perp\alpha_3}+
\right .
\\
\nonumber
\left . \left .
\epsilon^{(1)}_{\perp\mu_1}q_{\lam}
\varepsilon_{\lam\alpha_1\alpha_2\alpha_3} P_{\alpha_1}
q_{\alpha_2} \epsilon^{(2)}_{\perp\alpha_3}
+
\epsilon^{(2)}_{\perp\mu_1}q_{\lam}
\varepsilon_{\lam\alpha_1\alpha_2\alpha_3} P_{\alpha_1}
q_{\alpha_2} \epsilon^{(1)}_{\perp\alpha_3}\right)
\right]\ .
\ee
For the $4^{++}$ state, the operator reads as follows:
\be
4^{++}: \qquad
S^{(2)}_{\mu_1\mu_2}X^{(2)}_{\mu_3\mu_4}(q)+
S^{(2)}_{\mu_1\mu_3}X^{(2)}_{\mu_2\mu_4}(q)+
S^{(2)}_{\mu_1\mu_4}X^{(2)}_{\mu_2\mu_3}(q)
\ee
$$
-\frac23\left[
g^\perp_{\mu_1\mu_2}\left(
S^{(2)}_{\mu_3\lam}X^{(2)}_{\lam\mu_4}(q)+
S^{(2)}_{\mu_4\lam}X^{(2)}_{\lam\mu_3}(q)\right)+
g^\perp_{\mu_1\mu_3}\left(
S^{(2)}_{\mu_2\lam}X^{(2)}_{\lam\mu_4}(q)+
S^{(2)}_{\mu_4\lam}X^{(2)}_{\lam\mu_2}(q)\right)
\right .
\nonumber
$$
$$
\left .
+g^\perp_{\mu_1\mu_4}\left(
S^{(2)}_{\mu_2\lam}X^{(2)}_{\lam\mu_3}(q)+
S^{(2)}_{\mu_3\lam}X^{(2)}_{\lam\mu_2}(q)\right)\right] \ .
\nonumber
$$
General formula for the $J=L-1$ state,
after symmetrization over indices, reads:
$$
\varepsilon_{\mu_1\alpha_1\alpha_2\alpha_3} P_{\alpha_1}
X^{(L)}_{\alpha_2\nu
\mu_2\mu_3\mu_4\cdots\mu_{L-1}}(q)S^{(2)}_{\alpha_3\nu}\to
\varepsilon_{\mu_1\alpha_1\alpha_2\alpha_3} P_{\alpha_1}
X^{(L)}_{\alpha_2\nu
\mu_2\mu_3\mu_4\cdots\mu_{L-1}}(q)S^{(2)}_{\alpha_3\nu}+
$$
$$
+\varepsilon_{\mu_2\alpha_1\alpha_2\alpha_3} P_{\alpha_1}
X^{(L)}_{\alpha_2\nu
\mu_1\mu_3\mu_4\cdots\mu_{L-1}}(q)S^{(2)}_{\alpha_3\nu}+\cdots+
\varepsilon_{\mu_{L-1}\alpha_1\alpha_2\alpha_3} P_{\alpha_1}
X^{(L)}_{\alpha_2 \nu
\mu_1\mu_2\mu_3\cdots\mu_{L-2}}(q)S^{(2)}_{\alpha_3\nu}-
$$
$$
-\frac2{2L-1}\left(g^\perp_{\mu_1\mu_2}
\varepsilon_{\lam\alpha_1\alpha_2\alpha_3} P_{\alpha_1}
X^{(L)}_{\alpha_2\nu
\lam\mu_3\mu_4\cdots\mu_{L-1}}(q)S^{(2)}_{\alpha_3\nu}+
g^\perp_{\mu_1\mu_3}
\varepsilon_{\lam\alpha_1\alpha_2\alpha_3} P_{\alpha_1}
X^{(L)}_{\alpha_2\nu
\lam\mu_2\mu_4\cdots\mu_{L-1}}(q)S^{(2)}_{\alpha_3\nu}+
\right .
$$
$$
\left .
+\cdots+
g^\perp_{\mu_{L-2}\mu_{L-1}}
\varepsilon_{\lam\alpha_1\alpha_2\alpha_3} P_{\alpha_1}
X^{(L)}_{\alpha_2\nu
\lam\mu_1\mu_2\cdots\mu_{L-1}}(q)S^{(2)}_{\alpha_3\nu}
\right) \ .
$$

General formula for the $J=L$ state has the form:
\be
S^{(2)}_{\mu_1\nu}X^{(L)}_{\nu\mu_2\mu_3\ldots\mu_L}(q)\to
\sum^L_{i=1}S^{(2)}_{\mu_i\nu}
X^{(L)}_{\nu\mu_1\ldots\mu_{i-1}\mu_{i+1}\ldots\mu_L}(q)
\ee
$$
-\frac{2}{2L-1} \sum^L_{i,j=1 \atop i<j}
g^\perp_{\mu_i\mu_j}S^{(2)}_{\lam\nu}
X^{(L)}_{\lam\nu\mu_1\ldots\mu_{i-1}\mu_{i+1}\ldots\mu_{j-1}\mu_{j+1}
\ldots\mu_L}(q)
$$
General formula for the $J=L+1$ state has the form:
\be
\varepsilon_{\mu_1\alpha_1\alpha_2\alpha_3} P_{\alpha_1}
X^{(L)}_{\alpha_2\nu
\mu_2\mu_3\mu_4\cdots\mu_{L-1}}(q)S^{(2)}_{\alpha_3\nu}\to
\sum^L_{i=1}\varepsilon_{\mu_i\alpha_1\alpha_2\alpha_3} P_{\alpha_1}
X^{(L)}_{\alpha_2\nu\mu_1\ldots\mu_{i-1}\mu_{i+1}\ldots\mu_{L-1}}(q)
S^{(2)}_{\alpha_3\nu}
\ee
$$
-\frac{2}{2L-1} \sum^L_{i,j=1 \atop i<j}
g^\perp_{\mu_i\mu_j}\varepsilon_{\lam\alpha_1\alpha_2\alpha_3}
P_{\alpha_1}
X^{(L)}_{\alpha_2\nu\lam\mu_1\ldots\mu_{i-1}\mu_{i+1}\ldots
\mu_{j-1}\mu_{j+1}\ldots\mu_{L-1}}(q)
S^{(2)}_{\alpha_3\nu}\ .
$$
General formula for the $J=L+2$ state has the form:
\be
S^{(2)}_{\mu_1\mu_2}X^{(L)}_{\mu_3\mu_4\ldots\mu_{L+2}}(q)\to
\sum^L_{i,j=1 \atop i<j}S^{(2)}_{\mu_i\mu_j}
X^{(L)}_{\mu_1\ldots\mu_{i-1}\mu_{i+1}\ldots\mu_{j-1}\mu_{j+1}\ldots\mu_{L+2}}(q)
\ee
$$
-\frac{2}{2L-1} \sum^L_{i,j=1 \atop i<j}
g^\perp_{\mu_i\mu_j}\sum^L_{k=1 \atop k\ne i\ne j}
S^{(2)}_{\mu_k\nu}
X^{(L)}_{\nu\mu_1\ldots\mu_{i-1}\mu_{i+1}\ldots\mu_{j-1}\mu_{j+1}
\ldots\mu_{k-1}\mu_{k+1}\ldots\mu_L}(q) \ .
$$

\subsection{Odd-state operators, $P=-$}

The odd-spin operator, with $S=1$, is determined as
\be
\epsilon^{(1)}_{\alpha}\epsilon^{(2)}_{\beta}P_\gamma
\varepsilon_{\alpha\beta\gamma\mu}\equiv
S^{(1)}_{\mu}\ ,
\label{s1}
\ee
where $P_\mu=q_{1\mu}+q_{2\mu}$.
This structure should be multiplied by $X^{(L)}$, with
odd $L$:
\be
S^{(1)}_\mu X^{(L)}_{\mu_1...\mu_L}(q)\ .
\ee

This operator provides $J=L\pm 1$. The operator
for $J=L-1$ reads:
\be
S^{(1)}_\mu X^{(J+1)}_{\mu\mu_1...\mu_J}(q)\ .
\ee
It works for the states
\be
0^{-+}(L=1), \quad 2^{-+}(L=3), \quad 4^{-+}(L=5),\; ...
\ee

Without symmetrization, the operator for $J=L+1$ reads:
\be
S^{(1)}_{\mu_1} X^{(J-1)}_{\mu_2\mu_3...\mu_J}(q)-
\frac 13  g^\perp _{\mu_1\mu_2}
S^{(1)}_\mu X^{(J-1)}_{\mu\mu_3...\mu_J}(q)
\ee
where $ g^\perp _{\mu_1\mu_2} $ is working in the space perpendicular to
$p=q_1+q_2$. Generally, one should symmetrize this expression as
follows:
\be
S^{(1)}_{\mu_1}X^{(J-1)}_{\mu_2\mu_3\ldots\mu_J}(q)\to
\sum^J_{i=1}S^{(1)}_{\mu_i}
X^{(J-1)}_{\mu_1\ldots\mu_{i-1}\mu_{i+1}\mu_{J-1}}(q)
\ee
$$
-\frac{2}{2J-3} \sum^J_{i,j=1 \atop i<j}
g^\perp_{\mu_i\mu_j}S^{(1)}_{\lam}
X^{(J-1)}_{\lam\mu_1\ldots\mu_{i-1}\mu_{i+1}\ldots\mu_{j-1}\mu_{j+1}
\ldots\mu_J}(q)\ ,
$$
that gives $J$ terms. This operator is written for the states
\be
2^{-+}(L=1), \quad 4^{-+}(L=3), \quad 6^{-+}(L=5),\; ...
\ee
The state with $J=L$ is not formed by the transverse polarised
photons.

\subsection{The example: reaction $\gamma\gamma \to 2^{++}$-resonance
$\to \pi^0\pi^0 $}

To demonstrate  the connection of the introduced
operators with standard helicity technique, consider as an example
 the transition $\gamma\gamma\to 2^{++}-
resonance\to\pi^{0}\pi^{0} $.
Using the momenta $q_1$, $q_2$ for photons and
$k_1$, $k_2$ for mesons, one has for the relative momenta
and photon polarization vectors
\be
q=\frac12(q_1-q_2)=(0,0,0,q_z),
\qquad k=\frac12(k_1-k_2) =(0,k_x,k_y,k_z),
\ee
$$
\epsilon=(0,\epsilon_x,\epsilon_y,0)=(0,\cos{\phi},
\sin{\phi},0).
$$
In the helicity basis
\be
\epsilon=(0,\epsilon_+,\epsilon_-,0),
\ee
$$
\epsilon_+=-\frac1{\sqrt{2}}(\epsilon_x+i\epsilon_y),\qquad
\epsilon_-=\frac1{\sqrt{2}}(\epsilon_x-i\epsilon_y) \ .
$$
The spin-dependent part of the amplitude with $H=0$ reads:
\be
\epsilon^{(1)}_{\alpha}\epsilon^{(2)}_{\beta}S^{(0)}_{\alpha\beta}
X^{(2)}_{\mu\nu}(q)X^{(2)}_{\mu\nu}(k)=
\frac{9}{4}q^{2}k^{2}(\cos^2 \theta-\frac13)
(\epsilon^{(1)}_{+}\epsilon^{(2)}_{+}+\epsilon^{(1)}_{-}\epsilon^{(2)}_{-}).
\ee
For $H=2$ one has:
\be
\epsilon^{(1)}_{\alpha}\epsilon^{(2)}_{\beta}S^{(2)}_{\alpha\beta\,\mu\nu}
X^{(2)}_{\mu\nu}(k).
\ee
Different components are written as follows:
\be
\epsilon^{(1)}_{+}\epsilon^{(2)}_{+}S^{(2)}_{+ + \mu\nu}
X^{(2)}_{\mu\nu}(k^\perp)=0\ ,
\ee
$$
\epsilon^{(1)}_{+}\epsilon^{(2)}_{-}S^{(2)}_{+ - \mu\nu}
X^{(2)}_{\mu\nu}(k)=\frac32k^2
(1+2\sin^2{\phi}-2i\sin{\phi}\cos{\phi})=
\frac32 k^2(1-2i\sin{\phi}\,e^{i\phi})\ ,
$$
$$
\epsilon^{(1)}_{-}\epsilon^{(2)}_{+}S^{(2)}_{- + \mu\nu}
X^{(2)}_{\mu\nu}(k)=\frac32k^2
(1+2\sin^2{\phi}+2i\sin{\phi}\cos{\phi})=
\frac32 k^2(1+2i\sin{\phi}\,e^{-i\phi})\ ,
$$
$$
\epsilon^{(1)}_{-}\epsilon^{(2)}_{-}S^{(2)}_{- - \mu\nu}
X^{(2)}_{\mu\nu}(k)=0.
$$

\section{Conclusion}

We present a set of formulae for the angular--momentum and spin
operators, which have been used in the study of the reactions $p\bar p
\to SS,SP,PP$ \cite{anis-1,anis-2} as well as the moment--operator
expansion for $\gamma\gamma$-states. At the present time, the combined
analysis of
partial-wave amplitudes for the processes $p\bar p\to hadrons$ and
$\gamma\gamma \to hadrons$ is carried out on the basis of Crystal
Barrel and L3 Collaboration data.

The angular--momentum and spin operator expansions are rather helpful 
for studying hadron form factors and radiative decay transitions.The 
recent applications  of the operator expansion technique to the form 
factor processes can be found in papers \cite{f-gg,AA,phi}.

\section*{Acknowledgement}
We are grateful to D.V. Bugg, L.G. Dakhno, L. Montanet,
V.A. Nikonov and B.S.  Zou for stimulating discussions. The work was
supported by the RFFI grants N 01-02-17861,
N 01-02-17152 and N 00-15-96610.


\begin{thebibliography}{0}

\bibitem{anis-1} A.V. Anisovich, C.A. Baker, C.J. Batty et al.,
Phys. Lett. {\bf B452}, 173 (1999);
{\bf B468}, 304 (1999); {\bf B468}, 309 (1999).
\bibitem{anis-2}
A.V. Anisovich, C.A. Baker, C.J. Batty et al., Phys.
Lett. {\bf B452}, 187 (1999); {\bf B452}, 173 (1999);
{\bf B452}, 180 (1999);
{\bf B500}, 222 (2001);
{\bf B507}, 23 (2001).
\bibitem{klempt}E. Klempt, {\it "Meson spectroscopy: glueballs, hybrids,
and $q$ anti-$q$ mesons"}, hep-ex/0101031 (2001).
\bibitem{montanet} L. Montanet,
Nucl. Phys. Proc. Suppl. {\bf 86}, 381 (2000).
\bibitem{petry} R. Ricken, M. Koll, D. Merten, B.Ch. Metsch,
R.H. Petry, Eur. Phys. J. {\bf A6}, 221 (2000).
\bibitem{AAS} A.V. Anisovich, V.V.Anisovich, and A.V. Sarantsev,
Phys. Rev. D {\bf 62}:051502 (2000).
\bibitem{Kienzle} M.N. Kienzle-Focacci,
{\it "Two-photon collisions at LEP"},
in {\it Proceedings of the VIIIth Blois Workshop}, Protvino, Russia, 28
June-2 July 1999, eds.  V.A.  Petrov, A.V.  Prokudin, World
Scientific, 2000.
\bibitem{Scheg} V.A. Schegelsky, {\it "$\gamma\gamma$ physics at LEP"},
Talk given in Open Session of HEP Division of PNPI: {\it On
the Eve of the XXI Century}, 25-29 December 2000.
\bibitem{AD} V.V. Anisovich and L.G. Dakhno, JETP {\bf 44}, 198
(1963);\\
V.V. Anisovich and A.A. Anselm, UFN {\bf 88}, 287 (1965), [Sov.
Phys. Uspekhi {\bf 88}, 177 (1966)].
\bibitem{Zemach} C. Zemach, Phys. Rev. {\bf 97}, B97 (1965);
{\bf 97}, B109 (1965).
\bibitem{NN} A.V. Anisovich and A.V. Sarantsev, Sov. J. Nucl.
Phys. {\bf 55}, 1200 (1992).
\bibitem{AKMS} V. V. Anisovich, M. N. Kobrinsky, D. I. Melikhov and
A. V. Sarantsev, Nucl. Phys. A {\bf 544}, 747 (1992).
\bibitem{N-Delta} V.V. Anisovich, D.V. Bugg and A.V. Sarantsev,
Nucl. Phys. {\bf A537}, 501 (1992).
\bibitem{gamma-d} A.V. Anisovich and V.A. Sadovnikova,
Sov. J. Nucl. Phys. {\bf 55}, 1483 (1992); Eur. Phys. J.
{\bf A2}, 199 (1998).
\bibitem{3pi} A.V. Anisovich, Phys. of Atomic Nuclei, {\bf 62},
1412 (1999).
\bibitem{exp-3pi} A.V. Anisovich,
{\it "Three-body dispersion-relation $N/D$
equations for the coupled decay channels
$\bar p p$ ($J^{PC}=0^{-+}$) $ \rightarrow \pi^0 \pi^0
\pi^0$, $\eta \pi^0 \pi^0$, $\eta \eta \pi^0$,  $\bar K K \pi^0$"},
Phys. Atom. Nuclei, in press.
\bibitem{Chung} S.U. Chung, Phys. Rev. {\bf D57}, 431 (1998).
\bibitem{f-gg}
A. V.Anisovich, V. V.Anisovich, D. V. Bugg and V.A. Nikonov,
Phys. Lett. B {\bf 456}, 80 (1999).
\bibitem{AA}
A. V. Anisovich and V. V. Anisovich, Phys. Lett. B {\bf 467},
289 (1999).
\bibitem{phi} A.V. Anisovich, V.V. Anisovich and V.A. Nikonov
{\it "Radiative decays of basic scalar, vector and tensor mesons
and determination of the P-wave $q\bar q$ multiplet"},
hep-ph/0108186 (2001).
\bibitem{Hasan} A. Hasan and D.V. Bugg,
Phys. Lett. {\bf B334} (1998) 215 (1994).

\end{thebibliography}
\end{document}